\documentclass[fleqn,twoside]{article}
\usepackage{espcrc2}

\usepackage{graphicx}
\usepackage[figuresright]{rotating}
\usepackage{amssymb,bm,braket,psfrag,colortbl,booktabs,latexsym}

\newcommand{\AmS}{{\protect\the\textfont2
  A\kern-.1667em\lower.5ex\hbox{M}\kern-.125emS}}
  
\newcommand{\be}{\begin{equation}}  
\newcommand{\ee}{\end{equation}}  
\newcommand{\bea}{\begin{eqnarray}}  
\newcommand{\eea}{\end{eqnarray}} 
\newcommand{\ff}{f\hspace{-0.4em}f} 
\newcommand{\nn}{\nonumber} 
\newcommand{\Tr}{\mathrm{Tr}}

\title{Top-Quark Pair Production Beyond Next-to-Leading Order}

\author{V.~Ahrens\address[Mainz]{Institut f\"ur Physik (THEP), Johannes Gutenberg-Universit\"at,
    D-55099 Mainz, Germany}, 
    A.~Ferroglia\thanks{Talk
given at {\em Loops and Legs in Quantum Field Theory
2010}, W\"orlitz, Germany, April 25-30, 2010.}
\addressmark[Mainz]
\address{Physics Department, New York City College of Technology, 300 Jay Street,
    Brooklyn NY 11201, USA}, 
    M.~ Neubert\addressmark[Mainz] \address{ Institut f\"ur Theoretische Physik, Ruprecht-Karls-Universit\"at Heidelberg, Philosophenweg 16, D-6912
Heidelberg, Germany},
    B.~D.~Pecjak\addressmark[Mainz], and L.~L.~Yang\addressmark[Mainz]}
       
\begin{document}

\begin{abstract} 

We report on  recent calculations of the differential cross section for
top-quark pair production at hadron colliders. The results are differential with
respect to the top-pair invariant mass and to the partonic scattering angle.
In these calculations, which were carried out by employing   soft-collinear effective
theory techniques, we resummed threshold logarithms up to 
next-to-next-to-leading logarithmic order. Starting from the differential cross
section, it is possible to obtain theoretical predictions for the invariant-mass
distribution and the total cross section. We
summarize here our results for these observables, and we compare them with the
results obtained from different calculational methods.   

\vspace{1pc}
\end{abstract}

\maketitle

\section{INTRODUCTION}

The study of top-quark properties and in particular the measurements related
to the production of top-quark pairs   play a crucial role in the
physics programs at the Tevatron and at the LHC.  The large number of top-quark
pair events which will be detected at the LHC will turn the study of these
observables into precision physics.
Precise measurements require  equally precise theoretical predictions
for the quantities which are measured. Currently, complete next-to-leading order
(NLO) QCD calculations are available for  the total top-quark pair-production
cross section
\cite{Nason:1987xz,Beenakker:1988bq,Beenakker:1990maa,Czakon:2008ii}, as well as
for differential distributions 
\cite{Nason:1989zy,Mangano:1991jk,Frixione:1995fj}. The predictions
obtained from these calculations show theoretical uncertainties larger than $10
\, \%$ for both Tevatron and LHC center of mass energies.
The calculation of the NNLO corrections for the production of top-quark pairs
would be extremely useful. However, at present the two-loop virtual corrections which are needed to obtain complete NNLO calculations are known  only in part \cite{Czakon:2008zk,Bonciani:2008az,Bonciani:2009nb}. 
It is possible to improve the current NLO predictions by employing threshold resummation methods. For the total top-quark pair production cross section, complete next-to-next-to-leading logarithmic (NNLL)
expression were obtained very recently 
\cite{Beneke:2009rj,Czakon:2009zw,Beneke:2009ye}.
For what concerns the distributions differential with respect to the 
top-quark pair invariant mass and scattering angle, a NNLL resummation scheme was presented in \cite{Ahrens:2010zv}, while approximate NNLO formulas were first published in 
\cite{Ahrens:2009uz}. Both calculations were made possible by the evaluation of the needed two-loop anomalous-dimensions matrices \cite{Ferroglia:2009ep,Ferroglia:2009ii}.
In this work we summarize and comment the results reported in \cite{Ahrens:2010zv,Ahrens:2009uz}.


\section{DIFFERENTIAL CROSS SECTION}
We consider the process
\begin{equation}
N_1 (P_1) \,\,  N_2 (P_2) \to t(p_3) \,\,  \bar{t}(p_4) \,\, X(p_X) \, , \label{eq:process}
\end{equation}
where $N_1, N_2$ indicate either two protons (LHC case) or a proton and an antiproton
(Tevatron case), while $X$ indicates an inclusive hadronic final state. The doubly-differential cross section for the process in Eq.~(\ref{eq:process}) can be written as
\bea
 \label{eq:genfact}
  \frac{d^2\sigma}{dMd\cos\theta} &=& \frac{8\pi\beta_t}{3sM} \sum_{i,j} \int_\tau^1
  \frac{dz}{z} \, \ff_{ij}(\tau/z,\mu_f) \times \nn \\
  && \times
  \, C_{ij}(z,M,m_t,\cos\theta,\mu_f) \, ,
\eea
where $M$ is the pair invariant mass, $\theta$ is the scattering angle in the partonic frame, $m_t$ is the top quark mass,  $\ff_{ij}$ is the luminosity for the partonic initial state $ij$,  $s$ ($\hat{s}$)
is the hadronic (partonic) center of mass energy squared, and finally 
\be
\label{betatdef}
  z = \frac{M^2}{\hat{s}} \, , \quad \tau = \frac{M^2}{s} \, , \quad \beta_t =
  \sqrt{1-\frac{4m_t^2}{M^2}} \, .
\ee
The functions $C_{ij}$ in~(\ref{eq:genfact}) are the hard-scattering kernels
which describe the partonic process.
We define the {\em partonic threshold} as the region of the phase space in which
$\hat{s} \to M^2$ ($z \to 1$). The partonic threshold does not
coincide with the {\em production threshold} region, $\hat{s} \to 4 m_t^2$, 
which is usually considered in relation to the calculation of the total pair-production cross section. At the partonic threshold there is no phase space available for the emission of additional hard radiation in the final state, therefore higher order corrections in that region are dominated by virtual corrections and by the emission of soft gluons. 
Partonic processes which are not initiated by a quark-antiquark pair or by two gluons are also suppressed at the partonic threshold.
By using soft-collinear effective-theory methods, it is indeed possible to prove that in the $z \to 1$ limit the hard-scattering kernels can be written as \cite{Ahrens:2010zv}
\bea
  C_{ij}(z,M) &=& \Tr \Bigl[ \bm{H}_{ij}(M)  \bm{S}_{ij}(\sqrt{\hat{s}}(1-z)) \Bigr] \nn \\ &&+ \mathcal{O}(1-z) \, , \label{eq:Cfac}
\eea
where $ij \in \{q \bar{q}, gg\}$; the hard functions $\bm{H}$ and the soft functions $\bm{S}$ are matrices in color space. 
(All of  the objects in Eq.~(\ref{eq:Cfac}) also depend on the arguments $m_t,\cos\theta$, and on the factorization scale $\mu_f$.)
The hard functions are related to virtual corrections, while the soft functions originate from the real emission of soft gluons. At $n$-th order in perturbation theory, the soft functions involve plus distributions of the form
\be \label{eq:plus}
 \left[ \frac{\ln^m(1-z)}{1-z} \right]_+ ; \quad m = 0,\ldots,2n-1 \, .
\ee
In \cite{Ahrens:2009uz} we combined the explicit calculation of the hard and soft functions up to NLO with the information that can be extracted from the two-loop anomalous dimension matrices which appear in the renormalization group equations (RGE) satisfied by $\bm{H}$ and $\bm{S}$ \cite{Ferroglia:2009ep,Ferroglia:2009ii}; in this way we obtained approximate NNLO formulas for the differential cross section in (\ref{eq:genfact}). Such formulas include the analytic expression of the  coefficients multiplying the plus distributions  in the NNLO hard-scattering kernels, the scale dependent terms multiplying the delta function of argument $(1-z)$, and a specific set of subleading terms associated with the plus distributions\footnote{The latter originates from the fact that the soft functions are more naturally written in terms of the distributions
\begin{displaymath}
\left[ \frac{1}{1-z}\ln^m\left(\frac{M^2 (1-z)^2}{\mu^2 z} \right) \right]_+ \,, \!\!\!
\end{displaymath}
rather then in terms of the plus distributions in~\ref{eq:plus}; see \cite{Ahrens:2010zv}.}.
In \cite{Ahrens:2010zv} we studied the resummation of  the  distributions in
Eq.~(\ref{eq:plus}) up to NNLL directly in momentum space.  (The same procedure
was already successfully applied to other processes
\cite{Becher:2006nr,Becher:2006mr,Becher:2007ty,Ahrens:2008qu,Ahrens:2008nc,Becher:2009th}.)
This was done by solving the RGE satisfied by $\bm{H}$ and $\bm{S}$. The
explicit formulas for the NNLL resummed hard-scattering kernels can be found in
\cite{Ahrens:2010zv}, here we just want to remind the reader that they involve
two extra scales on top of the factorization scale: These are  the hard scale
$\mu_h$ and the soft scale $\mu_s$ which characterize the hard and soft
functions, respectively. At the moment of evaluating the resummed formulas
numerically, $\mu_s$, $\mu_h$, and $\mu_f$ are chosen independently from one
another. 
Both the approximate NNLO formulas and  the NNLL resummed formulas improve the
fixed order calculations of the hard-scattering kernels  at NLO  because  they
include higher order corrections originating from the partonic threshold region.
Does the partonic threshold region provide a numerically dominant contribution
to the cross section in Eq.~(\ref{eq:genfact})? Since for phenomenologically
interesting applications $\tau \lesssim 0.3$, at first sight this does not seem
to be the case. However, if the luminosities $\ff$ become small sufficiently
fast for $z \to \tau$, the integrals in Eq.~(\ref{eq:genfact}) will still be
dominated by the  region where $z \sim 1$. The latter effect goes under the
name of {\em dynamical threshold enhancement}. By comparing the exact NLO
invariant mass distribution with  the approximate NLO invariant mass
distributions obtained by retaining only the terms which are singular in the
threshold region (and the subleading terms associated to them), it is possible
to show that a dynamical threshold enhancement does take place in the 
production of top-quark pairs at the Tevatron and at the LHC
\cite{Ahrens:2010zv,Ahrens:2009uz}.

\section{PHENOMENOLOGY}

To obtain the best possible predictions, we matched the NNLL resummed formulas
to the NLO calculations for the invariant mass distribution and  the  total cross section. Also the approximate NNLO corrections
which we obtained are always added to the exact NLO results.
In the NNLL+NLO calculations, the numerical values of the scales are chosen in order to minimize the dependence of the NNLL corrections on $\mu_s, \mu_h$, and $\mu_f$;
in particular $\mu_f$ and $\mu_h$ are set equal to $M$, while $\mu_s$ 
is typically chosen to be in the interval $[M/4, M/10]$ (see\cite{Ahrens:2010zv} for details). For the same reasons, when dealing with approximate NNLO corrections, we set $\mu_f = M$. 
In general, once the NNLL or the approximate NNLO corrections are taken into account,  the residual scale uncertainty on the top-quark pair observables is smaller than  or comparable to the luminosity uncertainty.

\subsection{Invariant Mass Distribution}

\begin{figure}[t]
\begin{center}
\begin{tabular}{c}
\psfrag{x}[][][1][90]{$d\sigma/dM$ [fb/GeV]}
\psfrag{y}[]{$M$ [GeV]}
\psfrag{z}[][][0.85]{$\sqrt{s}=1.96$\,TeV}
\includegraphics[width=0.45\textwidth]{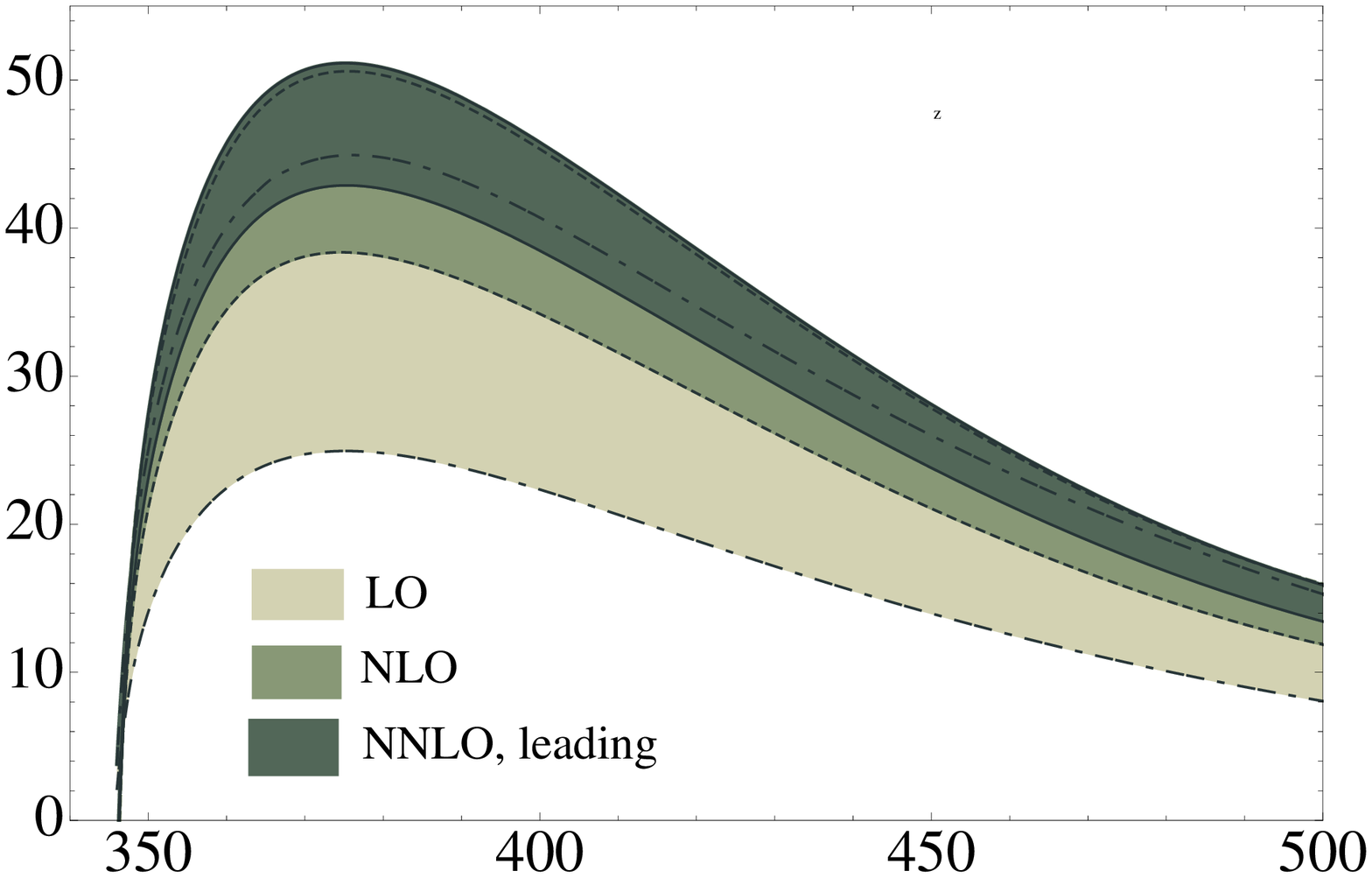}
\\[2mm]
\psfrag{x}[][][1][90]{$d\sigma/dM$ [fb/GeV]}
\psfrag{y}[]{$M$ [GeV]}
\psfrag{z}[][][0.85]{$\sqrt{s}=1.96$\,TeV}
\includegraphics[width=0.45\textwidth]{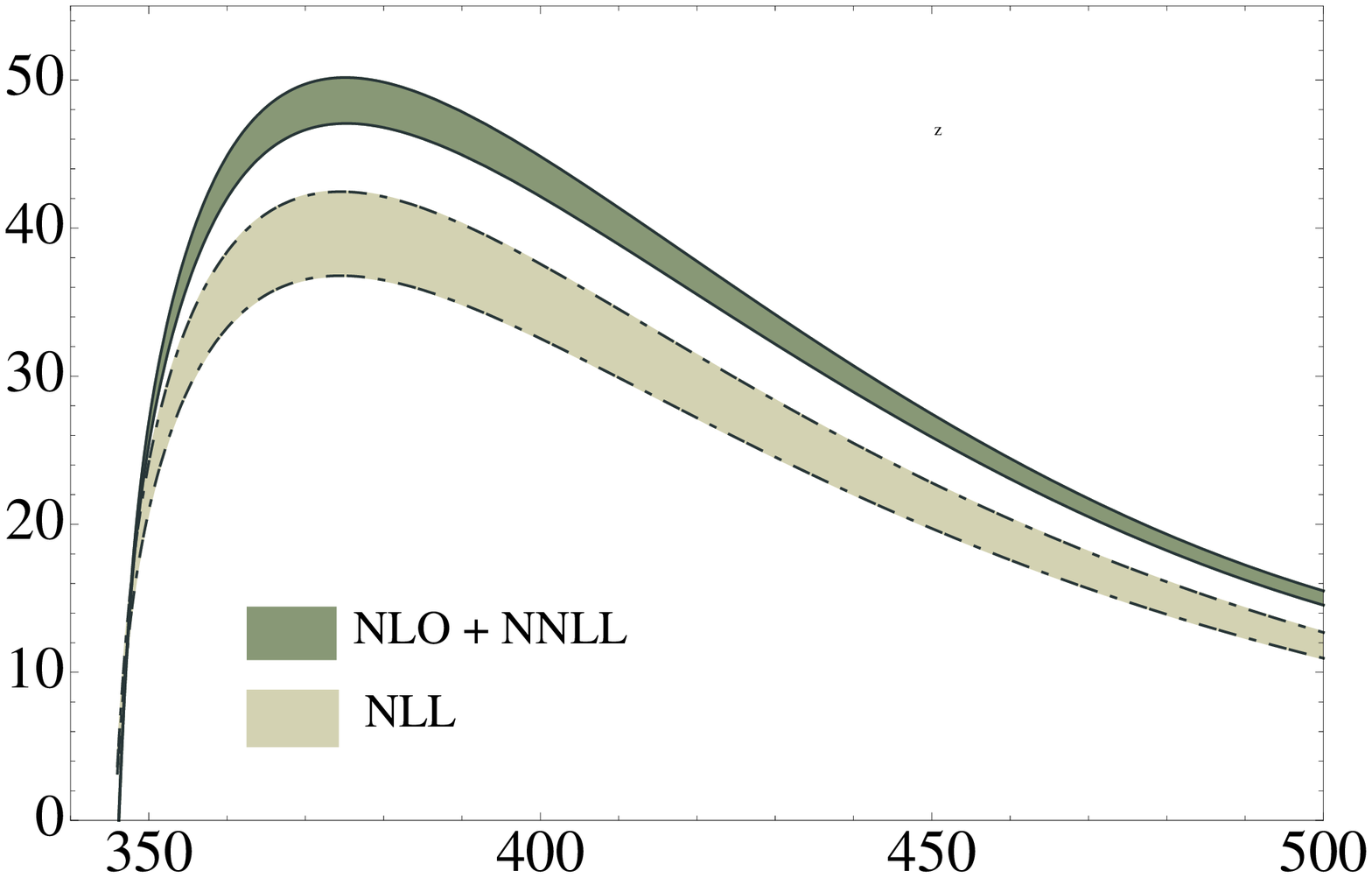}
\end{tabular}
\end{center}
\vspace{-11mm}
\caption{\label{fig:SpectrumPDFS} 
Top panel: Fixed-order predictions for the invariant mass spectrum at LO (light), NLO (darker), and approximate NNLO (dark bands) for the Tevatron. Bottom panel: Corresponding predictions at NLL (light) and NLO+NNLL (darker bands) in resummed perturbation theory.}
\vspace*{-8mm}
\end{figure}
\begin{figure}
\begin{center}
\psfrag{x}[][][1]{$M$ [GeV]}
\psfrag{y}[][][1][90]{$d\sigma/dM$ [fb/GeV]}
\psfrag{z}[][][0.85]{$\sqrt{s}=1.96$\,TeV}
\includegraphics[width=0.45\textwidth]{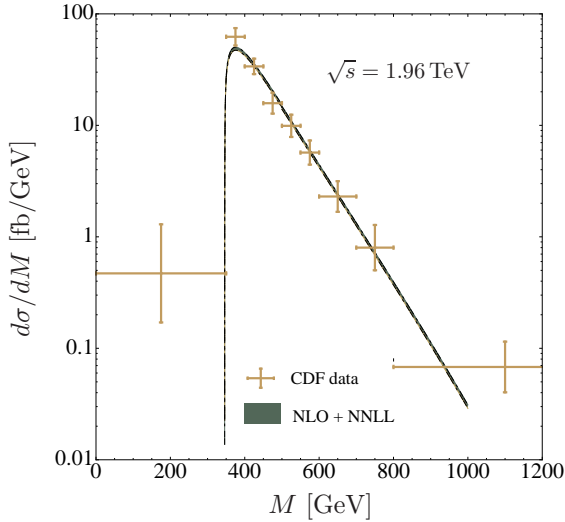} 
\end{center}
\vspace{-1.2cm}
\caption{\label{fig:CDF-compare} 
Comparison of the RG-improved predictions for the invariant mass spectrum with CDF data \cite{Aaltonen:2009iz}.}
\end{figure}

The invariant mass distribution can be obtained by integrating the 
doubly differential distribution of Eq.~(\ref{eq:genfact}) in the range $-1 < \cos \theta <1$.
The exact NLO results needed for the matching with the NNLL resummed and approximate NNLO corrections can be obtained by using the partonic Monte Carlo MCFM \cite{Campbell:2000bg}.
The results of the calculation of the invariant mass distributions at the
Tevatron  are shown in Fig.~\ref{fig:SpectrumPDFS}. (Similar plots for the LHC
case are available in \cite{Ahrens:2010zv}.) All the bands in
Fig.~\ref{fig:SpectrumPDFS} have been obtained by using MSTW2008 PDFs \cite{Martin:2009bu}.
In the top panel, the LO band was obtained by employing LO PDFs, the NLO band by
using NLO PDFs, etc. In the bottom panel, the NLL band was obtained by employing
NLO PDFs and the NNLL band by employing NNLO PDFs. 
The width of  the bands
reflects the scale uncertainty of the distributions. 
As expected, the width of the bands is significantly smaller in the lower panel, which shows the resummed results. The partonic threshold region corrections accounted for in our formulas  become more and more dominant as the value of $M$ increases.
 The CDF collaboration  released data for the invariant mass distribution in \cite{Aaltonen:2009iz}. We compared the data with our NLO+NNLL calculation and we found a good agreement, especially at large values of $M$ (see Fig.~\ref{fig:CDF-compare}).
%
%
It is possible to rewrite the invariant mass distribution in terms of a distribution with respect to the variable $\beta_t$ in Eq.~(\ref{betatdef}).
The $\beta_t$-distributions show clear  peaks for $\beta_t \sim 0.6-0.8$ at both the Tevatron and at the LHC. We observe that $\beta_t$ is always smaller than $\beta = \sqrt{1-4 m_t^2/\hat{s}}$. Therefore, values of the total cross section which are obtained by integrating the $\beta \to 0$ limit of the partonic cross section should be taken with a certain degree of caution, as we will illustrate further in the Section~\ref{totCS}.

\subsection{Total Cross Section \label{totCS}}

The total top-quark pair-production cross section can be calculated by
integrating the invariant mass distribution in the range $2 m_t < M < \sqrt{s}$.
An analytic NLO result for the total cross section can be found in
\cite{Czakon:2008ii}. In order to match our NNLL resummed results with the 
NLO cross section in \cite{Czakon:2008ii}, we chose $\mu_f =  400 \, \mbox{GeV}$, which is representative of  the values of $M$ at which the invariant mass distribution peaks. Our results for the total cross section are summarized  in Table~\ref{table:tot}.
\begin{table*}[htb]
\caption{Results for the total cross section in pb.
The first error refers to the perturbative uncertainties associated with scale variations, the second to PDF uncertainties. The factorization scale is fixed at $\mu_f =  400 \, \mbox{GeV}$.}
\label{table:tot}
\newcommand{\m}{\hphantom{$-$}}
\newcommand{\cc}[1]{\multicolumn{1}{c}{#1}}
\renewcommand{\tabcolsep}{2pc} 
\renewcommand{\arraystretch}{1.2} 
\begin{tabular}{llll}
\hline
$\mu_f =  400$ GeV           & \cc{Tevatron} & \cc{LHC ($7$ TeV)} &  \cc{LHC ($14$ TeV)} \\
\hline
$\sigma_{\mbox{{\tiny NLO}}}$       & \m$5.79^{+0.79+0.33}_{-0.80-0.22}$ & \m$133^{+21+7}_{-19-7}$ & \m$761^{+105+26}_{-101-27}$  \\
$\sigma_{\mbox{{\tiny NLO, leading}}}$       & \m$5.49^{+0.78+0.33}_{-0.78-0.20}$ & \m$134^{+16+7}_{-17-7}$ & \m$761^{+64+25}_{-75-26}$  \\ 
$\sigma_{\mbox{{\tiny NLO+NNLL}}}$       & \m$6.30^{+0.19+0.31}_{-0.19-0.23}$ & \m$149^{+7+8}_{-7-8}$ & \m$821^{+40+24}_{-42-31}$  \\
\hline
\end{tabular}
\end{table*}
In the second line of Table~\ref{table:tot} we report the results obtained by
retaining only the leading singular terms in the threshold limit $z \to 1$ in 
the expression of the NLO hard scattering kernels. By comparing those results 
with the ones for the exact NLO cross section (first line in
Table~\ref{table:tot}) it is immediate to see that the neglected  terms, which are subleading in $(1-z)$, contribute only
a few percent of the total cross section at NLO. This does not
represent a proof of the fact that the threshold expansion works well also at
higher orders in perturbation theory, however, it is quite reassuring.  
Our best predictions for the total cross section are obtained by matching the NLO calculations with the resummed formulas at NNLL and can be found in the last line of  Table~\ref{table:tot}. The resummation has a relatively small effect on the cross section  ($10-15 \%$ enhancement), but it reduces significantly the scale dependence of the predictions.
\begin{table*}[htb]
\caption{Results for the total cross section in pb.
The first error refers to the perturbative uncertainties associated with scale variations, the second to PDF uncertainties. The factorization scale is fixed at $\mu_f =  173.1 \, \mbox{GeV}$.}
\label{table:totmt}
\newcommand{\m}{\hphantom{$-$}}
\newcommand{\cc}[1]{\multicolumn{1}{c}{#1}}
\renewcommand{\tabcolsep}{2pc} 
\renewcommand{\arraystretch}{1.2} 
\begin{tabular}{llll}
\hline
$\mu_f =  173.1$ GeV           & \cc{Tevatron} & \cc{LHC ($7$ TeV)} &  \cc{LHC ($14$ TeV)} \\
\hline
$\sigma_{\mbox{{\tiny NLO}}}$       & \m$6.72^{+0.36+0.37}_{-0.76-0.24}$ & \m$159^{+20+8}_{-21-7}$ & \m$889^{+107+31}_{-106-32}$  \\
$\sigma_{\mbox{{\tiny NLO, leading}}}$       & \m$6.42^{+0.42+0.35}_{-0.76-0.23}$ & \m$152^{+7+8}_{-15-8}$ & \m$835^{+18+29}_{-60-30}$  \\ 
$\sigma_{\mbox{{\tiny NLO+NNLL}}}$       & \m$6.48^{+0.17+0.32}_{-0.21-0.25}$ & \m$146^{+7+8}_{-7-8}$ & \m$813^{+50+30}_{-36-35}$  \\
\hline
\end{tabular}
\end{table*}
The enhancement of the cross section due to resummation can be mimicked in the NLO calculation by
choosing a smaller factorization scale. The values obtained by choosing $\mu_f =
m_t$ are shown in Table~\ref{table:totmt}. While the resummed predictions are quite similar to the ones listed in Table~\ref{table:tot}, the NLO predictions in Table~\ref{table:totmt} are significantly higher than the ones in Table~\ref{table:tot}. 

\subsection{The Small $\bm{\beta}$-Expansion}

\begin{figure}[t]
\begin{center}
\begin{tabular}{c}
\psfrag{t}[][][0.75]{$q \bar{q}$-channel}
\psfrag{y}[][][1][90]{$\alpha_s$ correction [pb]}
\psfrag{x}[][][1]{$\beta$}
\psfrag{z}[][][0.75]{$\sqrt{s} = 1.96$ TeV}
\includegraphics[width=0.40\textwidth]{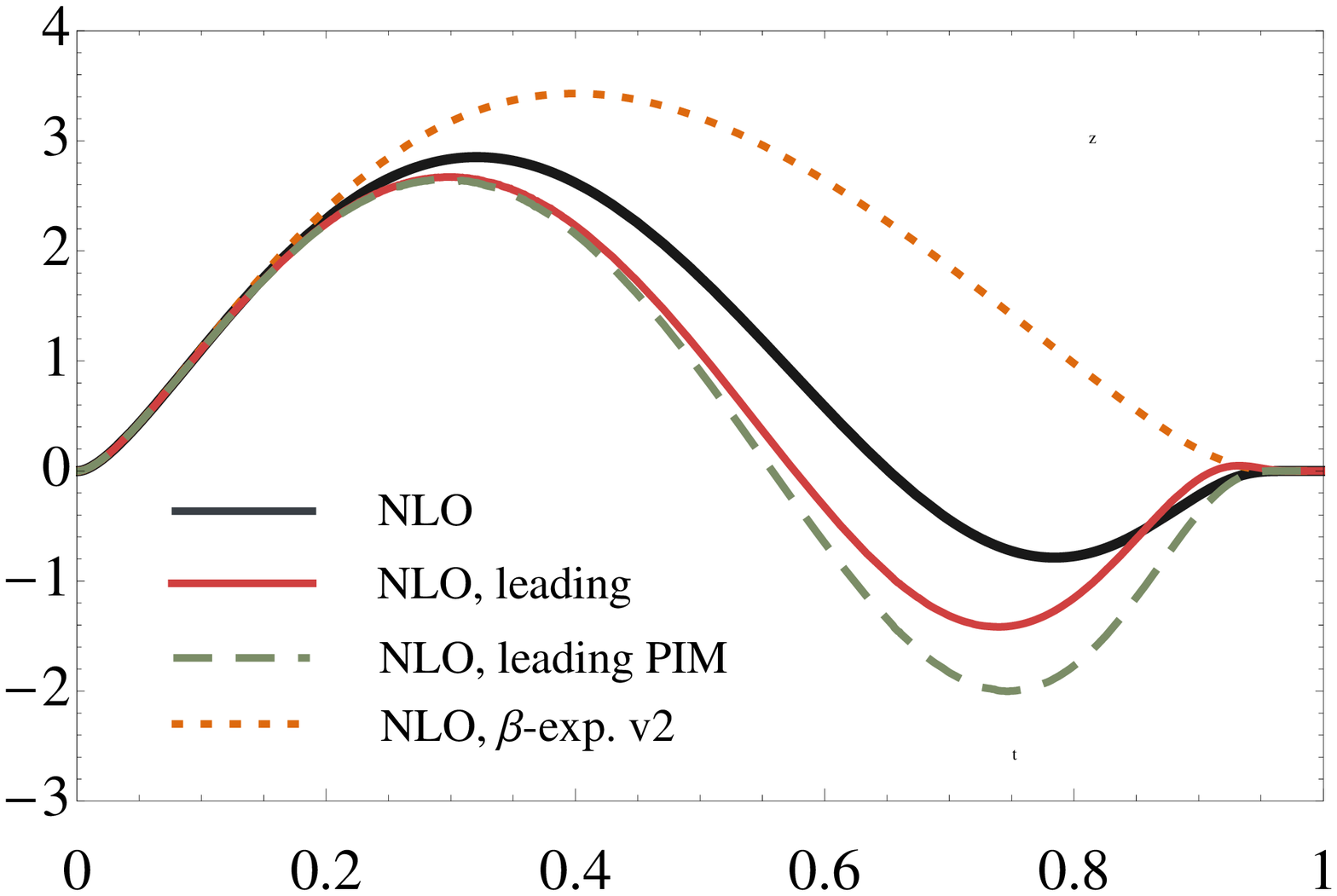}
\\[3mm]
\psfrag{t}[][][0.75]{$gg$-channel}
\psfrag{y}[][][1][90]{$\alpha_s$ correction [pb]}
\psfrag{x}[][][1]{$\beta$}
\psfrag{z}[][][0.75]{$\sqrt{s} = 1.96$ TeV}
\includegraphics[width=0.41\textwidth]{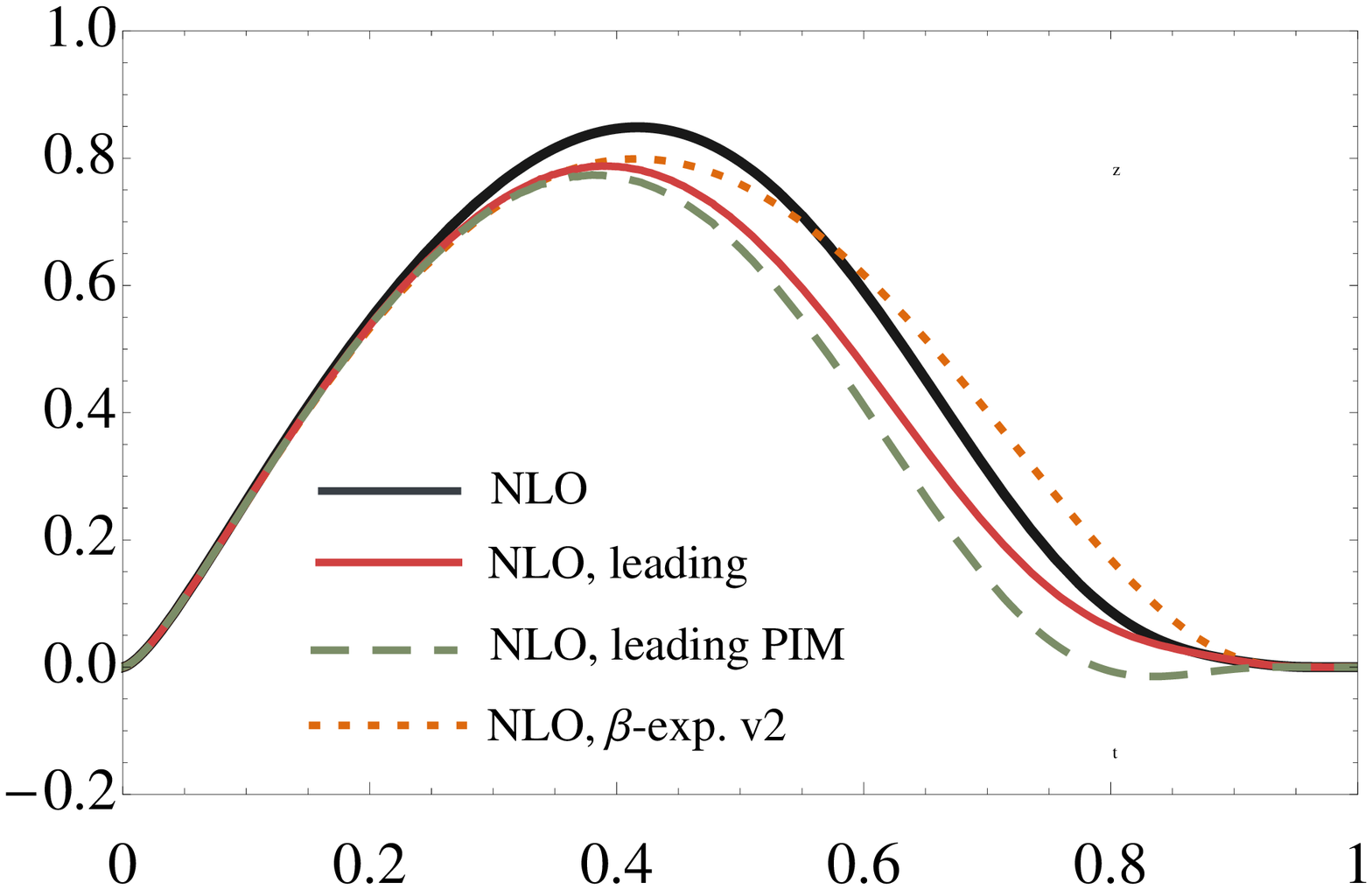}
\end{tabular}
\end{center}
\vspace{-12mm}
\caption{\label{fig:betatev} 
The ${\mathcal O}(\alpha_s)$ corrections to $d \sigma /d \beta$   at the Tevatron. Here $\mu_f = m_t$.}
\end{figure}

Recently, approximate NNLO calculations of the total partonic cross section
which are valid in the production threshold limit $\beta \to 0$  became
available \cite{Moch:2008qy,Langenfeld:2009wd,Beneke:2009ye}.
These approximate formulas include logarithmic terms  due to the emission of soft gluons and terms enhanced by Coulomb singularities.
By multiplying those expressions by the luminosity functions and subsequently integrating over $ 4 m_t^2 < \hat{s} < s$ it is possible to obtain values for the total cross section. 
The leading singular contributions in the $\beta \to 0$ limit do not coincide with those arising in the $z \to 1$ limit. However, by expanding our approximate NNLO formulas for the scattering kernels, which are valid in the partonic threshold region,  in the $\hat{s} \to 4 m_t^2$ limit and 
subsequently integrating over the scattering angle, it is possible to recover
the results of \cite{Beneke:2009ye} up to potential gluon terms.
%
\begin{figure}[t]
\begin{center}
\begin{tabular}{c}
\psfrag{t}[][][0.75]{$q \bar{q}$-channel}
\psfrag{y}[][][1][90]{$\alpha_s$ correction [pb]}
\psfrag{x}[][][1]{$\beta$}
\psfrag{z}[][][0.75]{$\sqrt{s} = 14$ TeV}
\includegraphics[width=0.41\textwidth]{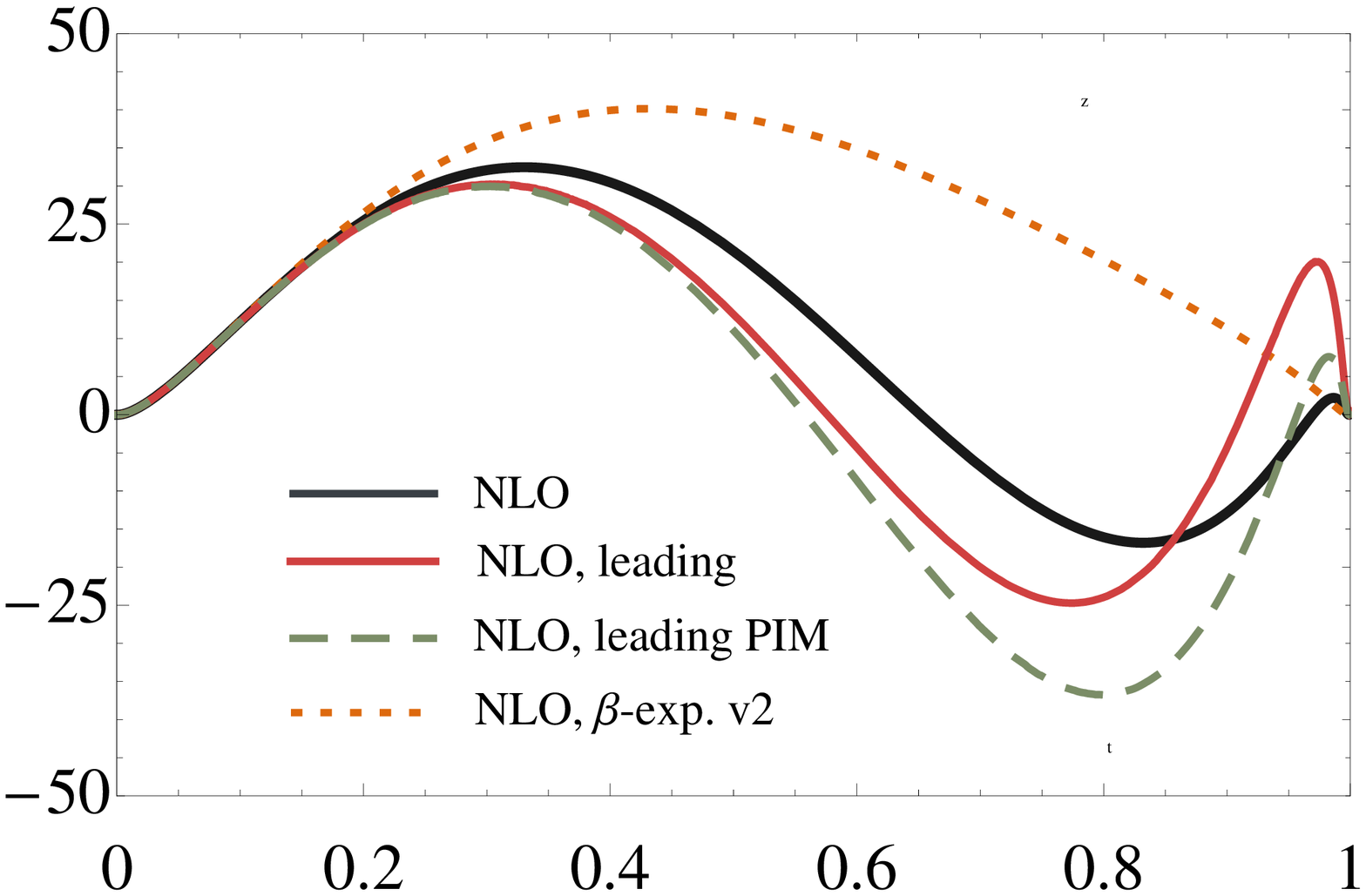}
\\[3mm]
\psfrag{t}[][][0.75]{$gg$-channel}
\psfrag{y}[][][1][90]{$\alpha_s$ correction [pb]}
\psfrag{x}[][][1]{$\beta$}
\psfrag{z}[][][0.75]{$\sqrt{s} = 14$ TeV}
\includegraphics[width=0.41\textwidth]{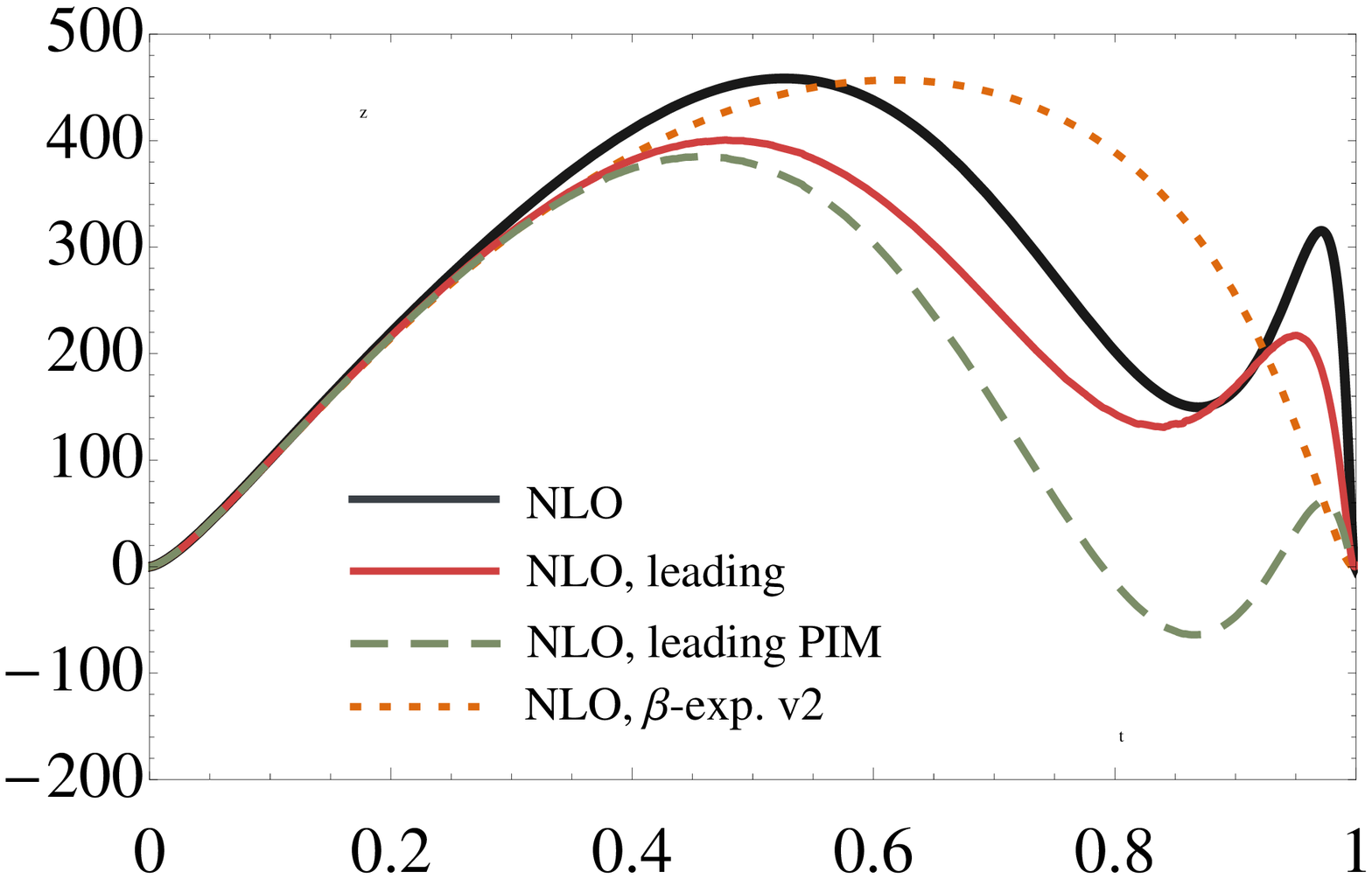} 
\end{tabular}
\end{center}
\vspace{-12mm}
\caption{\label{fig:betalhc} 
The ${\mathcal O}(\alpha_s)$ corrections to $d \sigma /d \beta$  at the LHC. Here $\mu_f = m_t$.}
\end{figure}
%
It is interesting to compare the predictions for the total cross section obtained
by integrating the formulas which are valid in the $\beta \to 0$ limit with the
ones obtained by integrating the  formulas which are valid in the $z \to 1$
limit.  This analysis can be carried out at NLO where exact results are available.
We consider then the product of the luminosity and of
the ${\mathcal O}(\alpha_s)$ corrections to the  partonic cross section as a
function of $\beta$. Those quantities are shown
in Figs.~\ref{fig:betatev} (Tevatron case) and  \ref{fig:betalhc} (LHC case). In
the figures, the solid black lines represent the exact NLO result, the  solid
red lines are our approximate NLO results (NLO leading), the dashed lines correspond
to the  approximate NLO
results  in which we drop all of the subleading terms (NLO leading
PIM), and finally the dotted lines indicate the  approximate NLO formulas in the $\beta
\to 0$ limit (NLO $\beta$-exp, see \cite{Ahrens:2010zv} for further details). 
The contribution of each of  these corrections to the total cross section is given by the area below the corresponding curve (the maximal value of $\beta$ allowed is $\sim 0.98$ at the Tevatron and $\sim 0.99$ at the LHC).
As expected, all the curves become indistinguishable as $\beta \to 0$.
However, the NLO leading curves reproduce the shape of the exact results also for $\beta \gg 0$, while the the NLO $\beta$-exp curves do not. 
In particular, the NLO $\beta$-exp overestimates the contribution to the cross section in the quark-antiquark channel at both the Tevatron and the LHC.
The NLO leading curves underestimate the corrections in the gluon channel; however, the good agreement between the area under the exact NLO curve and 
the area under the $\beta$-exp curve at the LHC looks rather accidental.
Finally, by comparing the NLO leading curves to the NLO leading
PIM curves, one sees that the subleading terms which are kept in the former sensibly improve the agreement with the exact NLO results.

\end{document}